\documentclass[12pt]{article}
\usepackage{amsmath}
\usepackage{amssymb}
\usepackage{slashed}
\usepackage[square,numbers,sectionbib,compress]{natbib}

\newcommand{\eps}{\varepsilon}
\newcommand{\Msb}{$\overline{\text{MS}}$}

\usepackage{pslatex}
\usepackage[latin1]{inputenc}
\usepackage[T1]{fontenc}

\usepackage[symbol]{footmisc}

\begin{document}

\numberwithin{equation}{section}

\begin{titlepage}
\noindent
DESY-22-099 \hfill June 2022\\
\vspace{0.6cm}
\begin{center}
{\LARGE \bf 
    Off-forward anomalous dimensions in the leading-$n_f$ limit\footnote[1]{Presented at Loops and Legs in Quantum Field Theory 2022.}}\\ 
\vspace{1.4cm}
\large
S.~Van Thurenhout$^{\, a}$ and S. Moch$^{\, a}$\\
\vspace{1.4cm}
\normalsize
{\it $^a$II.~Institute for Theoretical Physics, Hamburg University\\
\vspace{0.1cm}
D-22761 Hamburg, Germany}\\
\vspace{1.4cm}
{\large \bf Abstract}
\vspace{-0.2cm}
\end{center}
We review the computation of off-forward anomalous dimensions in the limit of a large number of quark flavors $n_f$. The method is based on a consistency relation the anomalous dimensions have to obey, which is a direct consequence of the renormalization structure of the operators in the chiral limit. In addition, we present a way to generate the anomalous dimensions to all orders in the strong coupling $\alpha_s$. This is based on exact conformal symmetry at the Wilson-Fisher critical point and provides an extension of previous calculations of this type for the forward anomalous dimensions.
\vspace*{0.3cm}
\end{titlepage}

\section{Introduction}
\vspace{-0.3cm}
While remarkable progress has been made in the last few decades concerning our knowledge of hadronic structure, a number of open questions still remains. An example is the proton spin puzzle \cite{Aidala:2012mv,Leader:2013jra,Deur:2018roz,Ji:2020ena}, which represents our ignorance of how the spins and angular momenta of quarks and gluons conspire to make the proton a spin-1/2 fermion. At the experimental level, one can gain insight into the hadronic structure by scattering electrons off protons at high energies. It is instructive to consider both inclusive processes, like inclusive deep-inelastic scattering (DIS), and exclusive ones, like deeply-virtual Compton scattering (DVCS), as they provide access to different properties of partonic interactions and distributions within hadrons. Experimental studies were performed e.g.~by {\sc HERA} \cite{Abramowicz:2015mha,Accardi:2016ndt} in the past, and will continue e.g.~with the planned Electron Ion Collider \cite{Boer:2011fh,AbdulKhalek:2021gbh}. On the theoretical side, the hadronic structure can be analyzed by studying matrix elements of composite operators, as these are related to parton distributions. Well-known examples are the standard parton distribution functions (PDFs) for inclusive processes and generalized parton distribution functions (GPDs) for exclusive ones. These distributions are universal quantities, in the sense that they do not depend on the specific process under consideration. Unfortunately, as the matrix elements are to be taken with respect to hadronic states, they can not be calculated using perturbation theory. Hence they have to be fitted to experimental data or studied using non-perturbative methods like lattice QCD, see e.g.~\cite{Ji:2020ena,Alexandrou:2020sml,Wang:2021vqy,Alexandrou:2021bbo,Scapellato:2022mai,Alexandrou:2022dtc,Alexandrou:2020zbe,Alexandrou:2021lyf} for recent progress. However, the scale dependence of the distributions can be determined in perturbation theory as an expansion in the strong coupling $\alpha_s$, as it is controlled by the anomalous dimensions of the operators which define them. These anomalous dimensions are computed by renormalizing the partonic matrix elements of the operators.

Our focus will be on exclusive processes, for which the operator anomalous dimensions are determined by renormalizing non-forward partonic operator matrix elements (OMEs), i.e.~the initial and final state partons have different momenta. This implies that the operators can mix under renormalization with total derivative ones, for which a basis has to be chosen. This then leads to an anomalous dimension matrix (ADM). While the diagonal elements of the ADM simply correspond to the forward anomalous dimensions, which are well-known from the study of inclusive processes \cite{Gross:1973ju,Floratos:1977au,Shifman:1980dk,Baldracchini:1981,Artru:1989zv,Gracey:1994nn,Hayashigaki:1997dn,Kumano:1997qp,Vogelsang:1997ak,Blumlein:2001ca,Gracey:2003mr,Moch:2004pa,Ablinger:2010ty,Velizhanin:2011es,Velizhanin:2012nm,Velizhanin:2014fua,Ruijl:2016pkm,Moch:2017uml,Herzog:2018kwj,Blumlein:2021enk}, the off-diagonal elements only appear for exclusive processes and require a separate calculation.

In the present article, we review the determination of the ADM for flavor non-singlet quark operators in the leading-$n_f$ limit. For this, we follow the method introduced in \cite{Moch:2021cdq}, in which it was shown that the entries of the ADM have to obey a consistency relation. This relation was derived by analyzing the renormalization structure of the operators in the chiral limit, and can be used to bootstrap the off-diagonal elements. Furthermore, we present a way to extract the leading-$n_f$ anomalous dimensions to all orders in perturbation theory. The method is based on exact conformal symmetry near the Wilson-Fisher fixed point and extends previous calculations of the same type in forward kinematics \cite{Gracey:1994nn,Gracey:2003mr}.

The article is organized as follows. First we introduce the operators and their evolution under scale variations. Next we briefly summarize the method to compute the off-forward anomalous dimensions in the leading-$n_f$ limit. We then apply this method to three types of operators, which are distinguished by their Dirac structure. The next section provides a way to generate all-order expressions for the elements of the ADM, and we finish with concluding remarks in Section \ref{sec:conclusion}.
\vspace{-0.3cm}
\section{Theoretical framework}
\vspace{-0.3cm}
We consider the renormalization of flavor non-singlet quark operators, which are defined as
\begin{equation}
\label{eq:operators}
    \mathcal{O} = \mathcal{S}\overline{\psi}\lambda^{\alpha}\Gamma D_{\nu_1}\dots D_{\nu_N} \psi.
\end{equation}
Here $D_{\mu}=\partial_{\mu}-ig_sA_{\mu}$ is the QCD covariant derivative and $\lambda^{\alpha}$ the generators of the SU($n_f$) flavor group. As we are interested in the leading-twist operators only, we symmetrize the Lorentz indices and subtract the traces, which is denoted by $\mathcal{S}$. Depending on the Dirac structure, generically represented by $\Gamma$ here, the operators in Eq.(\ref{eq:operators}) describe different physical phenomena. Three cases which are relevant for phenomenology are
\begin{itemize}
    \item $\Gamma=\gamma_{\mu}$ corresponding to Wilson operators,
    \item $\Gamma=\sigma_{\mu\nu}\equiv \frac{1}{2}[\gamma_{\mu},\gamma_{\nu}]$ for transversity operators and
    \item $\Gamma=\gamma_5\gamma_{\mu}$ in the case of polarization.
\end{itemize}
Depending on the kinematics of the process, hadronic matrix elements of the Wilson operators are related to PDFs or GPDs. The first characterize the longitudinal momentum and polarization carried by partons within hadrons, and are accessible e.g.~in inclusive DIS. The latter provide information on transverse distributions of the partons, and are accessible in exclusive processes like DVCS \cite{Ji:1996nm,Diehl:2003ny}. When considering a transversly polarized nucleon, the forward hadronic matrix elements of the transversity operator measure the difference in probabilities of finding a parton within the nucleon polarized in the same direction as the nucleon spin and finding a parton polarized in the opposite direction. They are relevant for hadronic processes like the polarized Drell-Yan process \cite{Artru:1989zv,Ralston:1979ys,Jaffe:1991kp,Jaffe:1991ra,Cortes:1991ja}. The corresponding non-forward matrix elements are accessible e.g.~in vector meson production and are related to transverse distribution amplitudes (DAs), which measure the parton distribution within the meson \cite{Diehl:2003ny,Mikhailov:2008my}. Finally, matrix elements of polarized operators give rise to polarized PDFs in forward kinematics and polarized GPDs and DAs in off-forward kinematics. Examples of relevant processes are longitudinally polarized DIS \cite{AsymmetryAnalysis:1999gsr} and $\gamma\gamma^{*}\rightarrow\pi^0$ transitions \cite{Jakob:1994hd,Kroll:1996jx,Radyushkin:1996tb,Musatov:1997pu}.

The scale dependence of the distributions is determined by the scale dependence of their defining operators, which is set by their anomalous dimension,
\begin{equation}
    \frac{\text{d}[\mathcal{O}]}{\text{d}\ln\mu^2} = \gamma[\mathcal{O}].
\end{equation}
The square brackets denote renormalized operators. Such operator anomalous dimensions can be calculated perturbatively in the strong coupling $a_s=\alpha_s/(4\pi)$,
\begin{equation}
    \gamma \equiv a_s \gamma^{(0)} +a_s^2 \gamma^{(1)} + \dots,
\end{equation}
by renormalizing the partonic matrix elements of the operators in Eq.(\ref{eq:operators}). For exclusive processes, these matrix elements are non-forward, i.e.~the initial and final state parton have different momenta. This means that the operators in Eq.(\ref{eq:operators}) mix under renormalization with total derivative ones. A basis then has to be selected for these additional operators. In this paper we exclusively use the total derivative basis \cite{Moch:2021cdq,Gracey:2009da,Gracey:2011zn,Gracey:2011zg,Geyer:1982fk,Blumlein:1999sc}, in which the operators are generically written as
\begin{equation}
\label{eq:Dbasis}
    \mathcal{O}_{p,q,r}^{\mathcal{D}} = (\Delta \cdot \partial)^{p} \Big\{(\Delta \cdot D)^{q}\overline{\psi}\, (\Delta\cdot \Gamma) 
    (\Delta \cdot D)^{r}\psi\Big\}.
\end{equation}
We have contracted the operators with a tensor $\Delta^{\mu_1}\dots\Delta^{\mu_{N+1}}$ whose components are lightlike, $\Delta^2=0$, to select the leading-twist contributions. The renormalization of the operators in Eq.(\ref{eq:operators}), including total derivatives, then follows
\begin{equation}
    \mathcal{O}_{k,0,N}^{\mathcal{D}} = \sum\limits_{j=0}^{N} Z_{N,N-j}^{\mathcal{D}}[\mathcal{O}_{k+j,0,N-j}^{\mathcal{D}}].
\end{equation}
In the chiral limit, operators of the form $\mathcal{O}_{k,N,0}^{\mathcal{D}}$ renormalize with exactly the same $Z$-factors, i.e.~
\begin{equation}
    \label{eq:leftD}
    \mathcal{O}_{k,N,0}^{\mathcal{D}} = \sum\limits_{j=0}^{N} Z_{N,N-j}^{\mathcal{D}}[\mathcal{O}_{k+j,N-j,0}^{\mathcal{D}}].
\end{equation}
The elements of the anomalous dimension matrix are then defined as
\begin{equation}
    \gamma_{N,k}^{\mathcal{D}} = -(Z_{N,\:j}^{\mathcal{D}})^{-1}\frac{\text{d}\:Z^{\mathcal{D}}_{j,k}}{\text{d}\ln\mu^2}.
\end{equation}
The diagonal elements, $k=N$, correspond to the forward anomalous dimensions. They determine the scale dependence of forward distributions through the DGLAP equation \cite{Gribov:1972ri,Altarelli:1977zs,Dokshitzer:1977sg}
\begin{equation}
    \frac{\text{d} f_{\text{NS}}(x,\mu^2)}{\text{d} \ln{\mu^2}} = \int_x^1 \frac{\text{d}y}{y} P_{\text{NS}}(y)f_{\text{NS}}\Big(\frac{x}{y},\mu^2\Big).
\end{equation}
Here $f_{\text{NS}}$ denotes a generic forward distribution and $P_{\text{NS}}$ the corresponding splitting function, which is related to the operator anomalous dimension by a Mellin transform
\begin{equation}
    \gamma_{N,N} = -\int_{0}^{1}\text{d}x \: x^{N}P_{\text{NS}}(x).
\end{equation}
Note that we can omit the superscript $\mathcal{D}$ here, as the diagonal elements do not depend on the basis chosen for total derivative operators. Similarly, the off-diagonal elements $\gamma_{N,k}^{\mathcal{D}}$ with $k\neq N$ determine the scale dependence of off-forward distributions through the ERBL equation \cite{Efremov:1978rn,Efremov:1979qk,Lepage:1979zb,Lepage:1980fj}
\begin{equation}
    \frac{\text{d}\phi(x,\mu^2)}{\text{d}\ln\mu^2} = \int_{0}^{1}\text{d}y\:V(x,y)\phi(y,\mu^2)
\end{equation}
with $\phi$ a generic non-forward distribution. The kernel $V(x,y)$ is related to the anomalous dimensions through \cite{Dittes:1988xz}
\begin{equation}
    \sum_{k=0}^N\gamma_{N,k}^{\mathcal{D}}\:y^k=-\int_{0}^{1}\text{d}x\:x^N\: V(x,y) .
\end{equation}
\vspace{-0.3cm}
\section{Method}
\vspace{-0.3cm}
It was shown in \cite{Moch:2021cdq} that, in the chiral limit, the anomalous dimensions have to obey the following consistency relation
\begin{equation}
    \label{eq:mainConj}
    \gamma_{N,k}^{\mathcal{D}} \,=\, 
    \binom{N}{k}\sum_{j=0}^{N-k}(-1)^j \binom{N-k}{j}\gamma_{j+k,j+k} 
    + \sum_{j=k}^N (-1)^k \binom{j}{k} \sum_{l=j+1}^N (-1)^l \binom{N}{l} \gamma_{l,j}^{\mathcal{D}}
    \, .
\end{equation}
This is valid to all orders in $a_s$ and allows one to reconstruct the off-diagonal part of the ADM based on the knowledge of the forward anomalous dimensions and the last column $\gamma_{N,0}^{\mathcal{D}}$. The latter was shown in \cite{Moch:2021cdq} to be related to the bare matrix elements of the operators in Eq.(\ref{eq:operators}).

We now focus our attention on the leading-$n_f$ limit, which leads to some important simplifications. First, only harmonic sums with positive indices appear in the expressions for the anomalous dimensions, which makes the sums in Eq.(\ref{eq:mainConj}) easier to handle\footnote[2]{The sums can generically be evaluated using concepts of symbolic summation, which are nicely implemented e.g.~in the {\sc Mathematica} package {\sc Sigma} \cite{Schneider2004}.}. Second, when the order in the strong coupling is increased by one, the maximum weight of the leading-$n_f$ term also just increases by one. It turns out that, because of these simplifications, the majority of terms in the $L$-loop anomalous dimensions can be predicted from the structure of the $(L-1$)-result. For example, when the $L$-loop anomalous dimension has a term proportional to $\frac{1}{N+2}$, the corresponding expression at the $(L+1)$-loop level will have a term $\frac{b_0}{(N+2)^2}$ with $b_0=\frac{2}{3}n_f$. A more complete set of such rules can be found in \cite{Moch:2021cdq}. Only a small number of unknowns then remains, which can be fixed by the consistency relation in Eq.(\ref{eq:mainConj}). 
\vspace{-0.3cm}
\section{Results}
\vspace{-0.3cm}
For the Wilson operators, the leading-$n_f$ anomalous dimensions were calculated in \cite{Moch:2021cdq} to order $a_s^4$ by combining a Feynman diagram calculation with the consistency relation in Eq.(\ref{eq:mainConj}). The $\zeta_N$-independent part, with $\zeta_N$ the Riemann-zeta function with argument $N$, of the five-loop expression was then determined using the method described above. However, the same method also applies to terms proportional to $\zeta_N$. As illustration, consider the four-loop $\zeta_3$ term, which was calculated in \cite{Moch:2021cdq} to be
\begin{equation}
    \gamma^{\mathcal{D}, (3)}_{N,k}\biggr \vert_{\zeta_3} \,=\, \frac{32}{27} n_f^3 C_F \zeta_3\Bigg(\frac{1}{N+2}-\frac{1}{N-k}\Bigg).
\end{equation}
We then expect the corresponding five-loop expression to be of the form
\begin{align}
\begin{split}
    \gamma^{\mathcal{D}, (4)}_{N,k}\biggr \vert_{\zeta_3} = &\:\: \frac{32}{27}n_f^4C_F\zeta_3\Bigg\{\frac{2}{3}\frac{1}{(N+2)^2}+\frac{2}{3}[S_1(N)-S_1(k)]\Bigg(\frac{1}{N+2}-\frac{1}{N-k}\Bigg)\Bigg\}\\&+n_f^4C_F\zeta_3\Bigg(\frac{a_1}{N+2}+\frac{a_2}{N+1}+\frac{a_3}{N-k}\Bigg)
\end{split}
\end{align}
with $a_i\in\mathbb{Q}$ a priori unknown. They can easily be determined however using the consistency relation Eq.(\ref{eq:mainConj}). With the five-loop expression for the forward anomalous dimension of \cite{Gracey:1994nn} we find
\begin{align}
\begin{split}
\label{eq:z3}
    \gamma^{\mathcal{D}, (4)}_{N,k}\biggr \vert_{\zeta_3} = &\:\: \frac{32}{27}n_f^4C_F\zeta_3\Bigg\{\frac{2}{3}\frac{1}{(N+2)^2}+\frac{2}{3}[S_1(N)-S_1(k)]\Bigg(\frac{1}{N+2}-\frac{1}{N-k}\Bigg)\\&-\frac{22}{9}\frac{1}{N+2}+\frac{4}{3}\frac{1}{N+1}+\frac{10}{9}\frac{1}{N-k}\Bigg\}.
\end{split}
\end{align}
Finally, the coefficient of the five-loop $\zeta_4$ term will be of minimal weight. This means that the consistency relation combined with the forward anomalous dimension of \cite{Gracey:1994nn} is enough to fix the corresponding off-diagonal part and we quickly find
\begin{equation}
\label{eq:z4}
    \gamma^{\mathcal{D}, (4)}_{N,k}\biggr \vert_{\zeta_4} = \:\: \frac{32}{27}n_f^4C_F\zeta_4\Bigg(\frac{1}{N+2}-\frac{1}{N-k}\Bigg).
\end{equation}
Eqs.(\ref{eq:z3}) and (\ref{eq:z4}) are new results.

Next, the one-loop anomalous dimensions for the transversity operators are extracted from a Feynman diagram calculation, combined with Eq.(\ref{eq:mainConj}). The higher-order results can then be determined using the method described above. This has been applied to order $a_s^4$ in \cite{VanThurenhout:2022nmx}.

Finally, in the leading-$n_f$ limit, the forward anomalous dimensions of the Wilson operators and the polarized ones coincide. The corresponding off-forward matrices will then also be the same, such that the results presented in \cite{Moch:2021cdq} and above can simply be reused.
\vspace{-0.3cm}
\section{All-order results}
\vspace{-0.3cm}
In \cite{Gracey:1994nn} and \cite{Gracey:2003mr} the all-order expressions for the Wilson and transversity forward anomalous dimensions in the leading-$n_f$ approximation were computed. The calculation relied on exact conformal symmetry at the Wilson-Fisher critical point \cite{Braun:2018mxm}, in which case propagators in the model simply have a power law structure. The anomalous dimensions calculated this way are then functions of the spacetime dimension $D$ and $n_f$. Here we extend this programme to the computation of the off-diagonal elements of the ADM. Defining
\begin{equation}
    \mu = \frac{D}{2} = 2-\eps
\end{equation}
and
\begin{equation}
    \eta = \frac{1}{n_f}\frac{(\mu-2)(2\mu-1)\Gamma(2\mu)}{\Gamma^2(\mu)\Gamma(\mu+1)\Gamma(2-\mu)}
\end{equation}
the general expression from which the anomalous dimensions can be extracted is\footnote{We thank A. Manashov for useful discussions on this subject.}
\begin{align}
    \begin{split}
    \label{eq:allorder}
        \gamma\mathcal{O}(z_1,z_2) =& \frac{\mu(\mu-1)}{2(\mu-2)(2\mu-1)}\eta\Bigg\{\int_{0}^{1}\text{d}\alpha\:\frac{\overline{\alpha}^{\mu-1}}{\alpha}(2[\mathcal{O}(z_1,z_2)]-[\mathcal{O}(z_{12}^{\alpha},z_2)]-[\mathcal{O}(z_1,z_{21}^{\alpha})])\\&-(\mu-\delta)^2\int_{0}^{1}\text{d}\alpha\int_{0}^{\overline{\alpha}}\text{d}\beta\:(1-\alpha-\beta)^{\mu-2}[\mathcal{O}(z_{12}^{\alpha},z_{21}^{\beta})]+\frac{\mu-1}{\mu}\mathcal[{O}(z_1,z_2)]\Bigg\}.
    \end{split}
\end{align}
Here $z_{12}^{\alpha} = z_1\overline{\alpha}+z_2\alpha$ and $\overline{\alpha} = 1-\alpha$. The parameter $\delta$ in Eq.(\ref{eq:allorder}) depends on the Dirac structure of the operators. Specifically we have $\delta=1$ for Wilson operators and $\delta=2$ for transversity ones. Note that the operators in Eq.(\ref{eq:allorder}) are non-local, i.e.~they depend on the two spacetime points $z_1$ and $z_2$. The $\eps$-parameter is understood to be the Wilson-Fisher one, i.e.~
\begin{equation}
    \eps\rightarrow\eps_{*} = -a_s\beta_0\biggr\vert_{n_f} = \frac{2}{3}n_fa_s.
\end{equation}
At present only the $O(a_s)$ term is needed. The forward anomalous dimensions are extracted from Eq.(\ref{eq:allorder}) by replacing 
\begin{equation}
  [ \mathcal{O}(z_1,z_2)] \rightarrow (z_1-z_2)^{N-1}.
\end{equation}
The resulting expressions agree with those in \cite{Gracey:1994nn,Gracey:2003mr}. To extract the off-forward anomalous dimensions, we use that the non-local operators act as generating functions for local ones \cite{Braun:2017cih},
\begin{equation}
\label{eq:lightcone}
    [\mathcal{O}(z_1,z_2)] \,=\, 
    \sum\limits_{m,k} \frac{z_1^m z_2^k}{m!\, k!}
    [\overline{\psi}(x) (\stackrel{\leftarrow}{D} \cdot \Delta)^k (\Delta\cdot\Gamma) (\Delta \cdot \stackrel{\rightarrow}{D})^m \psi(x)] \equiv \sum\limits_{m,k} \frac{z_1^m z_2^k}{m!\, k!}[\mathcal{O}_{k,m}]
    \, .
\end{equation}
Here $\Delta$ is an arbitrary lightlike vector, $\Delta^2=0$. Note that the local operators in the right-hand side can be written as
\begin{equation}
    [\mathcal{O}_{k,m}] = [\mathcal{O}^{\mathcal{D}}_{0,k,m}],
\end{equation}
cf.~Eq.(\ref{eq:Dbasis}). We now want to rewrite this in terms of operators in which covariant derivatives act only on the $\overline{\psi}$ field as we know how these renormalize, cf.~Eq.(\ref{eq:leftD}). This can be done by using the following operator identity\footnote{There is a small typo in \cite{Moch:2021cdq}; Eq.(2.25) there should be replaced by Eq.(\ref{eq:transBasis}) here.}
\begin{equation}
\label{eq:transBasis}
    \mathcal{O}_{0,N-k,k}^{\mathcal{D}} = (-1)^k\sum_{j=0}^{k}(-1)^j\binom{k}{j}\mathcal{O}_{j,N-j,0}^{\mathcal{D}}.
\end{equation}
In the following we omit the last index on the operators in the right-hand side, abbreviating
\begin{equation}
    \mathcal{O}_{j,N-j,0}^{\mathcal{D}} \equiv \mathcal{O}_{j,N-j}^{\mathcal{D}}.
\end{equation}
A simple calculation then leads to the following form for the non-local operator in Eq.(\ref{eq:lightcone})
\begin{equation}
    [\mathcal{O}(z_1,z_2)] = \sum_{k=0}^{N}\sum_{j=0}^{k}(-1)^{j+k}\binom{k}{j}\frac{z_1^{N-k}z_2^k}{k!(N-k)!}[\mathcal{O}_{j,N-j}^{\mathcal{D}}].
\end{equation}
After substituting into Eq.(\ref{eq:allorder}), the resulting integrals can be computed for fixed values of $N$. Next, we take the $N$-th derivative with respect to $z_1$ and take $z_1,z_2\rightarrow 0$. The expression then takes on the form
\begin{equation}
    \gamma\mathcal{O}(z_1,z_2) = \gamma_{N,N}[\mathcal{O}_{0,N}^{\mathcal{D}}]+\gamma_{N,N-1}^{\mathcal{D}}[\mathcal{O}_{1,N-1}^{\mathcal{D}}]+\gamma_{N,N-2}^{\mathcal{D}}[\mathcal{O}_{2,N-2}^{\mathcal{D}}]+\dots+\gamma_{N,0}^{\mathcal{D}}[\mathcal{O}_{N,0}^{\mathcal{D}}],
\end{equation}
from which the all-order expressions for $\gamma_{N,k}^{\mathcal{D}}$ with $k=0,1,\dots,N$ can be read off. We have checked that we reproduce the correct forward anomalous dimensions computed in \cite{Gracey:1994nn,Gracey:2003mr}. For the off-forward anomalous dimensions, we agree with our previous calculations presented above and in \cite{Moch:2021cdq,VanThurenhout:2022nmx}. As a non-trivial example, we present here the off-diagonal elements for the spin-four operators, i.e.~$\gamma_{3,k}^{\mathcal{D}}$. Defining
\begin{align}
    &\mathcal{F}(a_s,n_f) = -\frac{2^{3-4a_sn_f/3}}{9\pi^{3/2}n_f}\frac{\Gamma(5/2-2a_sn_f/3)\sin(2\pi a_sn_f/3)}{\Gamma(6-2a_sn_f/3)}
\end{align}
we find
\begin{align}
    &\gamma_{3,2}^{\mathcal{D}} = -4(a_sn_f-3)[36+a_sn_f(2a_sn_f-15)]\mathcal{F}(a_s,n_f),\\&
    \gamma_{3,1}^{\mathcal{D}} = 9[18+a_sn_f(2a_sn_f-11)]\mathcal{F}(a_s,n_f),\\&
    \gamma_{3,0}^{\mathcal{D}} = -24(a_sn_f-3)\mathcal{F}(a_s,n_f)
\end{align}
for the Wilson operators and
\begin{align}
    &\gamma_{3,2}^{\mathcal{D},T} = (3-a_sn_f)[135+8a_sn_f(a_sn_f-6)]\mathcal{F}(a_s,n_f),\\&
    \gamma_{3,1}^{\mathcal{D},T} = 9[15+a_sn_f(2a_sn_f-7)]\mathcal{F}(a_s,n_f),\\&
    \gamma_{3,0}^{\mathcal{D},T} = \frac{-3}{2a_sn_f-3}[45+4a_sn_f(4a_sn_f-9)]\mathcal{F}(a_s,n_f)
\end{align}
for the transversity ones.
\vspace{-0.3cm}
\section{Conclusion and outlook}
\label{sec:conclusion}
\vspace{-0.3cm}
We have reviewed the computation of the anomalous dimensions of flavor non-singlet quark operators, including mixing with total derivative ones, in the large-$n_f$ limit. This was done in the total derivative basis, in which the anomalous dimensions have to obey a consistency relation. The origin of this relation lies in the renormalization structure of the operators in the chiral limit. Combined with the simple functional form of the leading-$n_f$ expressions, this then allows one to recursively construct the anomalous dimension matrices order per order in perturbation theory.

We also presented a way to generate the all-order expressions for the leading-$n_f$ anomalous dimension matrices, based on exact conformal symmetry at the Wilson-Fisher fixed point. This extends previous calculations of the same type by J. Gracey for the forward anomalous dimensions.

It should be noted that the large-$n_f$ anomalous dimensions are, by themselves, not particularly useful for phenomenology. However, such results are still important, since (a) they do contribute to the full expressions in QCD and (b) they can teach us about the structure of the anomalous dimensions. Moreover, the leading-$n_f$ calculations can be regarded as proofs-of-concept for the method based on the consistency relation between off-forward anomalous dimensions. A possible continuation would then be to apply the method in the leading-color limit, which typically provides a good approximation to the full result and hence is phenomenologically relevant. Finally, the method used for computing the all-order expressions could also be generalized to obtain subleading-$n_f$ contributions. These aspects are left for future studies.

\subsection*{Acknowledgements}
The authors would like to thank J. Gracey and A. Manashov for useful discussions and comments on the manuscript.
This work has been supported by Deutsche Forschungsgemeinschaft (DFG) through the Research Unit FOR 2926, ``Next Generation pQCD for
Hadron Structure: Preparing for the EIC'', project number 40824754 and DFG grant $\text{MO~1801/4-1}$.

\providecommand{\href}[2]{#2}\begingroup\raggedright\endgroup


\end{document}